\documentclass[sigplan]{acmart}
\usepackage[ruled,vlined]{algorithm2e}
\usepackage{graphicx}
\usepackage[caption=false,font=footnotesize]{subfig}
\AtBeginDocument{%
  }

\begin{document}

\acmYear{2026}\copyrightyear{2026}
\setcopyright{cc}
\setcctype[4.0]{by}
\acmConference[EuroMLSys '26]{The 6th Workshop on Machine Learning and Systems}{April 27--30, 2026}{Edinburgh, Scotland Uk}
\acmBooktitle{The 6th Workshop on Machine Learning and Systems (EuroMLSys '26), April 27--30, 2026, Edinburgh, Scotland Uk}
\acmDOI{10.1145/3805621.3807625}
\acmISBN{979-8-4007-2605-7/26/04}

\title{Opinion Depolarization in Social Networks using GNNs}

\author{Konstantinos Mylonas}
\email{kmylonas@tuc.gr}
\affiliation{%
  \institution{Technical University of Crete}
  \city{Chania}
  \state{Crete}
  \country{Greece}
}

\author{Thrasyvoulos Spyropoulos}
\email{spyropoulos@tuc.gr}
\affiliation{%
  \institution{Technical University of Crete}
  \city{Chania}
  \state{Crete}
  \country{Greece}
}

\renewcommand{\shortauthors}{Mylonas et al.}

\begin{abstract}
Nowadays, social media is a primary ground for political debate and exchange of opinions. A significant body of research suggests that these platforms are highly polarized, often exhibiting echo chamber structures where users connect mainly with like-minded individuals and limit their exposure to diverse content. Existing work based on popular opinion diffusion models shows that convincing a subset of “key” users to adopt moderate opinions can reduce the polarization of the entire network. Building on this insight, we propose an efficient algorithm to identify a set of users whose moderation minimizes polarization. Unlike prior approaches that require iteratively solving a large analytical model, our method leverages Graph Neural Networks (GNNs) to significantly reduce decision time while achieving close-to-oracle selection accuracy.
\end{abstract}




\maketitle

\section{Introduction}
In today’s digital landscape, dominated by platforms such as Instagram, Facebook, and TikTok, users are exposed to an overwhelming volume of content \cite{bakshy2015exposure}. Yet, despite this abundance, their actual information diets are often narrow. Homophily drives users to connect with like-minded individuals, while psychological mechanisms such as \emph{biased assimilation} and \emph{selective exposure} further limit engagement with opposing viewpoints \cite{greitemeyer2009biased, frey1986recent}. Additionally, recommendation algorithms tend to prioritize agreeable content, reinforcing existing beliefs. Together, these factors contribute to polarization and the formation of ideologically homogeneous communities, commonly known as echo chambers.

Polarized networks with two echo chambers (e.g., conservatives versus democrats) have been extensively studied in the literature. Previous work \cite{matakos2017measuring} suggests that polarization can be mitigated by convincing a subset of users to adopt more moderate stances on contentious topics. In their seminal paper, Matakos, Terzi, and Tsaparas \cite{matakos2017measuring}  introduced the \textbf{ModerateExpressed} problem, which asks for a set of $K$ users who, if convinced to adopt more moderate opinions, would minimize the polarization of the network. Polarization is defined as a function of users’ opinions, modeled according to the Friedkin and Johnsen opinion framework \cite{friedkin1990social}.

To address the \textbf{ModerateExpressed} problem, they proposed a greedy algorithm referred to as \textbf{GreedyExt}. This algorithm identifies the $K$ nodes to moderate iteratively. At each round, it picks a single node, who if moderated, would lead to the most (additional) reduction in total polarization. While this is essentially a greedy algorithm, which is not necessarily optimal, it often leads to great results. 
Nevertheless, this selection process has considerable computational complexity. At each step of the algorithm, every user must be considered individually: their moderation is simulated, the resulting polarization is recomputed, and the user leading to the greatest polarization decrease is selected. Thus, for a graph with $n$ nodes, $n$ expensive evaluations are required per step, resulting in a total of $K \times n$ computationally expensive evaluations to select all $K$ nodes.

In this work, we revisit the \textbf{ModerateExpressed} problem and propose a more scalable solution. Specifically, we leverage machine learning to accelerate the selection process. We train a lightweight Graph Neural Network (GNN) to predict the polarization decrease associated with each node. In this way, the computationally expensive evaluations of \textbf{GreedyExt} are replaced with efficient forward-pass approximations. Our results demonstrate that this approach, which we refer to as \textbf{GNN-GreedyExt}, achieves polarization reduction comparable to \textbf{GreedyExt}, while reducing runtime by a factor of 16 on a Twitter retweet graph with $5000$ nodes. The performance gap widens as the network size increases.

The remainder of this paper is organized as follows. Section~\ref{sec:opinionmodelmoderateexpressed} introduces the Friedkin and Johnsen opinion model, formally defines the \textbf{ModerateExpressed} problem, and describes the baseline \textbf{GreedyExt} algorithm. Section~\ref{sec:3} presents our proposed \textbf{GNN-GreedyExt} method. Section~\ref{sec:4} provides a brief overview of the real-world datasets used in our study, while Section~\ref{sec:5} reports the experimental results on polarization reduction and runtime performance. Finally, Section~\ref{sec:6} concludes the paper and outlines directions for future work.

\section{Related Work}

A substantial body of literature addresses the problem of reducing polarization in social networks. Most existing approaches focus on \emph{how} to depolarize users, primarily through content-based or friendship-based recommendation mechanisms. Content-oriented methods aim to expose users to diverse viewpoints by adjusting the information they receive. For example, \cite{vendeville2022opening} propose a theoretical model of content diffusion and optimize the injection of curated recommendations to maximize newsfeed diversity via a constrained quadratic program.

Another line of work studies the impact of friend recommendation systems on echo chambers and polarization. \cite{cinus2022effect} analyze how such systems may, under certain conditions, reinforce rather than mitigate polarization, depending on the network structure. Similarly, \cite{tommasel2021want} leverage Graph Convolutional Networks (GCNs) and NLP techniques to model user representations and recommend diverse friendships across echo chambers. These works focus on designing mechanisms that promote exposure to opposing views.

In contrast to this extensive literature on \emph{how} to intervene, relatively fewer works examine \emph{whom} to target for intervention in order to achieve maximal depolarization. A notable exception is \cite{matakos2017measuring}, which formulates the problem of selecting a set of $K$ users whose moderation, under the Friedkin–Johnsen opinion model and a polarization metric, yields the largest decrease in overall network polarization.

Our work is directly motivated by this formulation. We address the same core optimization problem—identifying the most influential set of $K$ users to moderate—but propose a method designed to scale more effectively to large networks. Additionally, we provide a precise mathematical optimization formulation of the problem \textbf{ModerateExpressed}, which was previously described only at a conceptual level rather than as an explicit objective with formal constraints. By focusing on the strategic selection of users rather than the design of specific depolarization mechanisms, our approach complements existing intervention-based methods and provides a principled way to allocate limited intervention resources efficiently.


\section{Preliminaries and Problem Definition}
\label{sec:opinionmodelmoderateexpressed}

\subsection{Opinion Dynamics and Polarization Index}
We represent a social network as an undirected graph $G(V, E)$, where $V$ is the set of nodes and $E$ is the set of edges. If node $u$ is connected with node $v$ then the tuple $(u, v)\in E$ and $w_{uv}$ is a weight associated with the edge $(u,v)$. We assume the well-studied Friedkin-Johnsen (FJ) opinion formation model \cite{friedkin1990social}, where each node $i$ has two ``features''~\cite{biondi2023dynamics}. The \emph{external} (or \emph{expressed}) opinion $z_i$ essentially captures how the node is perceived by its peers, based e.g. on the polarity of content circulated by $i$. The \emph{internal} opinion $s_i$ captures the user's personal biases and deeply held beliefs. In the FJ model, the latter is assumed unchanged (at least over the time scales of interest) and can be perceived as the ``stubborness'' of that node. The expressed opinion, on the other hand, evolves over time, influenced by the peers's expressed opinion according to the following formula:.


\begin{center}
    \begin{equation}\label{eqn:FK}
        z_i = \frac{s_i + \sum_{j\in N(i)} w_{ij}z_j}{1 + \sum_{j\in N(i)} w_{ij}}
    \end{equation}
\end{center}

Both $s_i$ and $z_i$ take values in the range $[-1, +1]$ where $+1$ implies complete agreement while $-1$ complete disagreement with a topic (for example COVID19 vaccination).

It has been shown that if every person $i$ updates her opinion based on $\eqref{eqn:FK}$, then the expressed opinions converge to a unique opinion vector $\mathbf{z^{\mathrm{eq}}}$, where each component $z_i^{\mathrm{eq}}$ is the opinion of user $i$ at the equilibrium \cite{matakos2017measuring}. The steady state vector $\mathbf{z^{\mathrm{eq}}}$ can be computed either by running the evolution process until convergence or by solving the following linear system \cite{bindel2015bad}:

\begin{equation}\label{eqn:Equilibrium}
\mathbf{(L+I)z^{\mathrm{eq}} = s} \Leftrightarrow     \mathbf{z^{\mathrm{eq}} = (L+I)^{-1}s},
\end{equation}

where $L$ is the weighted Laplacian matrix defined as:

$$
\mathbf{L} = 
\begin{cases} 
    \sum_{j\in N(i)}w_{ij}, & \text{if } i=j \\
    -w_{ij} & \text{if } i \neq j \text{ and } (i,j)\in E\\
    0 & \text{otherwise}
\end{cases}
$$

and $\mathbf{I}$ is the identity matrix. $\mathbf{L + I}$ is invertible as a positive definite matrix. Notice that based on \eqref{eqn:Equilibrium} the steady state vector $\mathbf{z^{\mathrm{eq}}}$ depends only on the internal opinions $\mathbf{s}$ and the Laplacian matrix $\mathbf{L}$ of the underlying graph.

Without loss of generality and in order to facilitate comparison between our results and the results of \cite{matakos2017measuring}, we use the same polarization metric, known as \emph{polarization index}. Polarization index is defined as:

\begin{equation}\label{eqn:PolarizationIndex}
    \pi(\mathbf{z^{\mathrm{eq}}}) = \frac{||\mathbf{z^{\mathrm{eq}}}||_{2}^{2}}{|V|}
\end{equation}


In the edge case where $\mathbf{z^{\mathrm{eq}} = 0} $, all users express neutral opinion and thus the polarization index is minimized: $\pi(\mathbf{z^{\mathrm{eq}}}) = 0$. We divide by $|V|$ in order to make the index independent of the network's size. 


\subsection{ModerateExpressed Problem}
\label{2b}


In the introduction, we discussed the problem under consideration informally. This section presents a formal definition of the ModerateExpressed problem. To do so, we first introduce the necessary notation.

Assume we have a network and a set of nodes $T$ which we aim to moderate. We want to compute the new equilibrium $\mathbf{z^{\mathrm{eq}}_{\mathrm{new}}}$ after moderating (setting $z_i =0 \quad \forall i\in T $) the nodes in $T$. The most straightforward way to do this is to run the evolution process until convergence using \eqref{eqn:FK}\footnote{It is important to manually prohibit the moderated nodes to depart from zero while running the evolution process}.  

Alternatively, the new steady state vector $\mathbf{z^{\mathrm{eq}}_{\mathrm{new}}}$, after setting and fixing the external opinions $z_i =0,\ \forall i \in T$, can be computed through the following process, as explained in \cite{matakos2017measuring}:

\begin{enumerate}
    \item Set $s_i = 0, \quad \forall i \in T$
    \item Update the Laplacian Matrix $\mathbf{L}$ with a row of zeros at the $i$-th row, $\forall i \in T$
    \item Get the new $\mathbf{z^{\mathrm{eq}}}$ from \eqref{eqn:Equilibrium} using the new $\mathbf{s}$ and $\mathbf{L}$ in the equation.
\end{enumerate}

Let's assume the binary control vector $\mathbf{x}$, where $x_i = 0$ means that node i is moderated ($z_i = 0$), while $x_i=1$ means that node i is not moderated. Furthermore, let $\mathbf{D_x}$ be the diagonal matrix with the components of $\mathbf{x}$ on its main diagonal. Then based on the process that we just described, the new equilibrium $\mathbf{z^{\mathrm{eq}}_{\mathrm{new}}}$ can be computed as:

\begin{equation}
     \mathbf{z^{\mathrm{eq}}_{\mathrm{new}}} = (\mathbf{D_{x}}\mathbf{L + I})^{-1} \cdot (\mathbf{s \odot x})
     \label{eqn:z}
\end{equation}

where $\odot$ is the Hadamard product (element-wise multiplication) between $\mathbf{s}$ and $\mathbf{x}$. The matrix $\mathbf{D_{x}}$ is used to fill with zeros the rows of $\mathbf{L}$ dictated by the vector $\mathbf{x}$, while the $\mathbf{(s \odot x)}$ part of expression \eqref{eqn:z} is responsible to zero out the corresponding components of $\mathbf{s}$. As we can see $\mathbf{z^{\mathrm{eq}}_{\mathrm{new}}}$ is a function of $\mathbf{x}$ and thus, the polarization index $\pi$ is a function of $\mathbf{x}$. 

Finally, let:

\begin{equation}
    f(\mathbf{x}) = \frac{|| (\mathbf{D_{x}}\mathbf{L + I})^{-1} \cdot (\mathbf{s \odot x}) ||_2^2}{|V|}
    \label{eqn:f}
\end{equation}

Here $f$ is simply the polarization index $\pi$ of the new steady state vector, expressed as a function of $\mathbf{x}$

With that in mind we are now ready to define formally the problem \textbf{ModerateExpressed}:

\begin{center}
    \begin{align*}
        &\text{minimize}_{\mathbf{x}}\  f(\mathbf{x})\\
        &\text{subject to}\ \sum_{i=1}^{|V|} x_i = K
    \end{align*}
\end{center}

With this formulation, we are essentially looking for the binary vector $\mathbf{x}$ with exactly $K$ non zero components, such that the quantity $f\mathbf{(x)}$ in Equation \ref{eqn:f} is minimized. Simply put, the problem \textbf{ModerateExpressed} asks for a set of $K$ users who, when their expressed opinions are set to zero, the polarization of the network is minimized.

\subsection{Baseline Algorithm GreedyExt}

An intuitive difficulty of the above problem is deciding whether to moderate a user with an extreme opinion but few connections, or a user with a milder
opinion who is embedded in a highly polarized neighborhood. It is formally shown in ~\cite{matakos2017measuring} that the problem is NP-hard. To this end, the authors propose a Greedy iterative algorithm, coined GreedyExt, that works as follows:

\textbf{GreedyExt} starts with an empty set $T^0$. At each step $t$ the algorithm adds to the existing solution $T^{t-1}$ the node $v$ which when setting $z_v=0$ causes the largest decrease $\pi(\mathbf{z^{\mathrm{eq}}_{t-1}}) - \pi(\mathbf{z^{\mathrm{eq}}_{t}})$ in the polarization index. Here $\mathbf{z^{\mathrm{eq}}_{t}}$ is the steady state vector after setting $z_i=0\quad \forall i\in T^t $.

Algorithm~\ref{alg:greedyext} presents the pseudocode for \textbf{GreedyExt}. The inner loop constitutes the most computationally expensive component, as it is executed for every node and requires computing the new steady-state vector $\mathbf{z}_{\mathrm{new}}$. This can be obtained either by running the opinion dynamics process in \eqref{eqn:FK} until convergence or by applying the method described in Section~\ref{2b}. In our experiments, consistent with \cite{matakos2017measuring}, we adopt the latter approach. However, as noted in their work, both approaches become prohibitively slow for large-scale graphs.

\begin{algorithm}[t]
\small
\caption{GreedyExt}
\label{alg:greedyext}

\KwIn{Graph $G = (V,E)$, $\mathbf{s}$, $\mathbf{z}$, timesteps $K$}
\KwOut{Solution set $T$}

$T \gets \emptyset$\;

\For{$t \gets 1$ \KwTo $K$}{
    
    $\pi_{\min} \gets +\infty$\;
    
    \ForEach{$node \in V \setminus T$}{
        
        $z_{node} \gets 0$\;
        
        Compute steady-state vector $\mathbf{z_{new}}$\;
        
        $\pi \gets \dfrac{\|\mathbf{z_{new}}\|_2^2}{|V|}$\;
        
        \If{$\pi < \pi_{\min}$}{
            $\pi_{\min} \gets \pi$\;
            $v^* \gets node$\;
        }
        
        Reset opinions to their original state\;
    }
    
    $T \gets T \cup \{v^*\}$\;
    
    $z_{v^*} \gets 0$\;
    
    Compute steady-state vector $\mathbf{z_{new}}$\;
    
    Update $G$ with $\mathbf{z_{new}}$\;
    
    Isolate node $v^*$\;
}

\Return{$T$}

\end{algorithm}


The \textbf{ModerateExpressed} problem is NP-hard, and therefore \textbf{GreedyExt} does not guarantee an optimal solution. However, although a formal proof is still missing (we plan to address this in future work), greedy algorithms are often near-optimal when the underlying objective function is submodular \cite{krause2014submodular}. Despite its effectiveness in identifying a good solution, \textbf{GreedyExt} suffers from high computational complexity: evaluating the polarization index after moderating each node independently is computationally expensive. Specifically, solving the linear system (\ref{eqn:Equilibrium}) for \textbf{one} node involves the multiplication of an $n \times n$ matrix with an $n$-dimensional vector, which translates to $O(n^2)$, where $n=|V|$. This process is repeated for every node and thus the complexity becomes $O(n^3)$. Repeating $K$ times (because we are looking for $K$ nodes), leads to $O(Kn^3)$. 

Fortunately, the authors of \cite{matakos2017measuring} observed that for the specific opinion model, $\mathbf{z^{\mathrm{eq}}}$ can be computed more efficiently, bringing the complexity of \textbf{GreedyExt} down to $O(Kn^2)$. Despite a clear improvement in speed, this complexity can still be prohibitive even for moderate values of $K$ and $n$ especially, as we'll also show in our experiments. Furthermore, variations of the FK model \cite{biondi2023dynamics} (as well as other opinion evolution models) abound in related work, and there's no guarantee that such analytical shortcuts (as the one that brought the complexity down to $O(Kn^2)$) exist for these models.

\section{Proposed Solution}
\label{sec:3}

Given the aforementioned shortcomings of  \textbf{GreedyExt}, we introduce, in this section, a new algorithm \textbf{GNN-GreedyExt}, that leverages the use of a GNN to speed up the selection process.

\subsection{GNN-GreedyExt}

\begin{figure}[t]
\centerline{\includegraphics[width=\columnwidth]{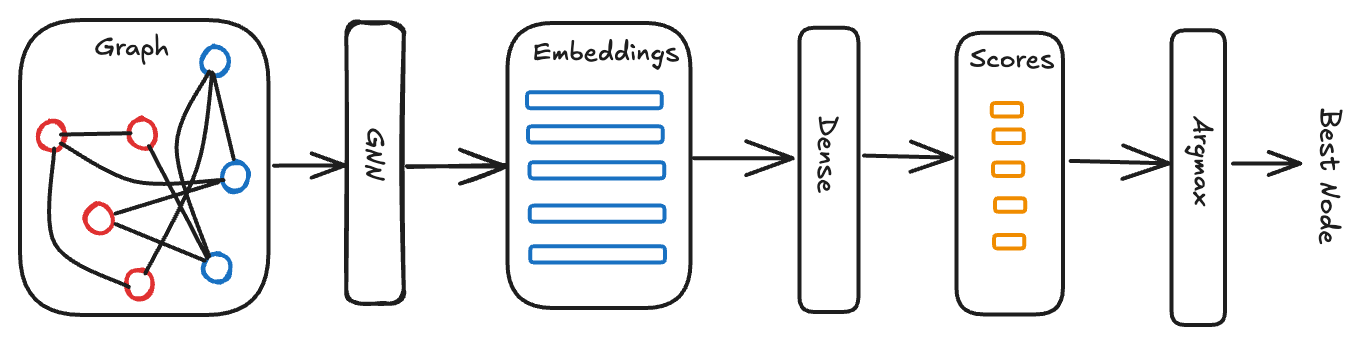}}
\caption{Graphical illustration of the pipeline that finds the best node at a timestep.}
\label{fig:pipeline}
\end{figure}

Similar to \textbf{GreedyExt}, \textbf{GNN-GreedyExt} is an iterative algorithm which runs for $K$ timesteps and adds nodes one by one to the solution set. Its main difference from \textbf{GreedyExt} is how the node is selected at each timestep. While \textbf{GreedyExt} computes the steady state vector $\mathbf{z^{\mathrm{eq}}}$ and the polarization decrease after moderating each node explicitly, our algorithm uses a GNN followed by a linear layer to approximate the polarization decrease for each node, without having to compute the steady state vector $\mathbf{z^{\mathrm{eq}}}$ at all. This way the expensive inner loop of Algorithm \ref{alg:greedyext} is replaced from cheap forward passes of a GNN. Algorithm \ref{alg:gnnalgorithm} contains the pseudocode for \textbf{GNN-GreedyExt} while a schematic illustration of the process for \textbf{one} timestep (out of $K$) is provided in Fig. \ref{fig:pipeline} for a more intuitive understanding. As shown in Fig. \ref{fig:pipeline}, for each timestep the graph is fed to a GNN which constructs one embedding per node. Those embeddings capture useful information about the node and its neighborhood, which is used from the subsequent linear layer in order to produce a vector of scores. The score $i$ serves as a proxy for the polarization decrease that would occur if node $i$ were moderated (i.e., if $z_i = 0$). The node with the highest score is then selected and added to the solution set. This process repeats for $K$ timesteps.

\begin{algorithm}[t]
\small
\caption{GNN-GreedyExt}
\label{alg:gnnalgorithm}

\KwIn{Graph $G = (V,E)$, $\mathbf{s}$, $\mathbf{z}$, timesteps $K$, GNN}
\KwOut{Solution set $T$}

$T \gets \emptyset$\;

\For{$t \gets 1$ \KwTo $K$}{
    
    $\textit{scores} \gets \mathrm{Predict}_{\mathrm{GNN}}(G)$\;
    
    $v^* \gets \arg\max_{v \in V \setminus T} \textit{scores}(v)$\;
    
    $T \gets T \cup \{v^*\}$\;
    
    $z_{v^*} \gets 0$\;
    
    Compute steady-state vector $\mathbf{z_{new}}$\;
    
    Update $G$ with $\mathbf{z_{new}}$\;
    
    Isolate node $v^*$\;
}

\Return{$T$}

\end{algorithm}

The \textbf{GNN-GreedyExt} algorithm relies on a GNN, which must be trained prior to deployment. To this end, suitable training data are required. In the following subsections, we describe the synthetic graphs used for training as well as the training configuration of our model.

\subsection{GNN Training}
We evaluate the performance of \textbf{GNN-GreedyExt} on the \textbf{ModerateExpressed} problem using both synthetic and real networks. Since real-world networks that include both a dual echo chamber structure and explicit node opinions are difficult to obtain, training the GNN directly on real data was not feasible. Instead, we generated synthetic graphs designed to capture key structural properties of real networks, such as the example shown in Fig.~\ref{fig:dcsbm}. Specifically, we construct artificial graphs with two echo chambers using a degree-corrected stochastic block model (DCSBM). This model was introduced in \cite{karrer2011stochastic}. The difference of this model with other stochastic block models lies in the degree distribution. In the DCSBM model the degree distribution of the graph follows a power law. The details of the DCSBM model can be found in the paper \cite{karrer2011stochastic}.

We construct 128 graphs with the DCSBM model, where each graph consists of 1000 nodes. The degrees are drawn from a power law distribution, which is common in social networks according to \cite{newman2010networks}. Next we assign internal opinions to nodes based on their community membership. Nodes $v$ in community 1 are set $s_v = +1$ while nodes $u$ in community 2 are set $s_u = -1$. Finally, $\mathbf{z^{\mathrm{eq}}}$ is computed and $z_i^{\mathrm{eq}}$ is attached to each node $i$. As a result every node consists of two features: $s_i$, and $z_i^{\mathrm{eq}}$.

Finally, in order to use these graphs for training, we need to compute the polarization decrease for each node $v$. Sometimes we refer to the polarization decrease as $Gain(v)$ which is formally defined as:

 $$Gain(v) = \pi(\mathbf{z^{\mathrm{eq}}_{old}}) -\pi(\mathbf{z^{\mathrm{eq}}_{new}}),$$
 
where $\mathbf{z^{\mathrm{eq}}_{old}}$ and $\mathbf{z^{\mathrm{eq}}_{new}}$ are the steady state vectors of external opinions before and after setting $z_v = 0$. The resulting graphs can be used for supervised training where the model learns to predict the gain after setting each node to zero. After the training is complete the model is ready to be used for inference in the \textbf{ModerateExpressed} problem.

\subsection{Model Architecture}

In principle, any type of GNN could be employed. In this work, we adopt the Graph Convolutional Network (GCN) \cite{kipf2016semi} due to its simplicity and effectiveness. Our architecture consists of two GCN layers, which aggregate information from neighbors up to two hops away, followed by a linear layer that outputs the predicted gain associated with moderating each node.  

Preliminary experiments indicated that more complex architectures did not provide noticeable improvements, likely because two-hop neighborhood information suffices to capture whether a node lies in a highly polarized region of the network and whether its moderation would result in a significant polarization decrease. The hidden representation dimension is set to 16.

\section{Real Datasets}
\label{sec:4}

We present results on three real-world networks exhibiting dual echo chambers with known community memberships. The experimental results are discussed in the next section.

\begin{table}[b]
\caption{Characteristics of the real networks considered}
\centering
\begin{tabular}{ | c | c | c | c | } 
  \hline
  Graph & $|V|$ & $|E|$ & $\pi$ \\ 
   \hline\hline
  Books & 105  & 441 & 0.107 \\ 
  \hline
  LiveJournal & 2766 & 24138 & 0.179 \\ 
  \hline
  WireTaping & 5000 & 201876 & 0.011 \\ 
  \hline
\end{tabular}
\label{table:1}
\end{table}

\paragraph{Political Books}
The Political Books dataset consists of political books sold on Amazon, where edges represent frequent co-purchasing. Each book is labeled as Conservative, Liberal, or Neutral. We map these labels to opinions as follows: Conservative $=+1$, Neutral $= 0$, and Liberal $= -1$. This mapping is chosen because precise numerical quantification of each node’s opinion is not available. An alternative would be to assign opinions by sampling values from the negative spectrum for Liberal nodes, the positive spectrum for Conservative nodes, and values around zero for Neutral nodes. While this would alter the resulting polarization of the network and potentially the set of nodes selected by the algorithms, the overall methodology remains unaffected.

Since our objective is to evaluate whether the proposed \textbf{GNN-GreedyExt} algorithm produces selections consistent with those of \textbf{GreedyExt}, the exact initialization of node opinions is not critical. In both cases, \textbf{GreedyExt} will apply its greedy selection process, and our method should exhibit comparable behavior. Therefore, without loss of generality, we adopt the $\{+1, 0, -1\}$ mapping, which is consistent with \cite{matakos2017measuring} and facilitates direct comparison with their results.

\begin{figure}[t]
    \centering
    \subfloat[DCSBM]{%
        \includegraphics[width=0.25\linewidth]{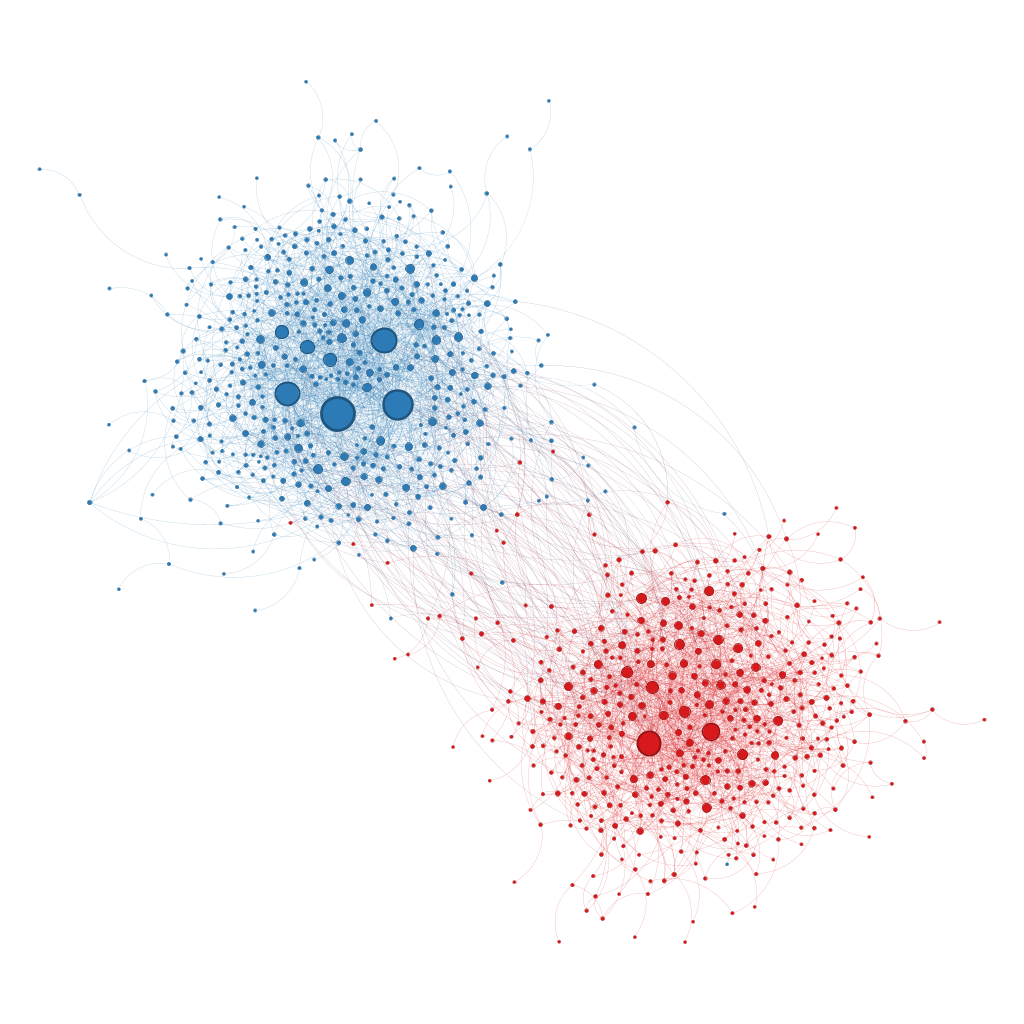}%
        \label{fig:dcsbm}}
    \hfill
    \subfloat[LiveJournal]{%
        \includegraphics[width=0.25\linewidth]{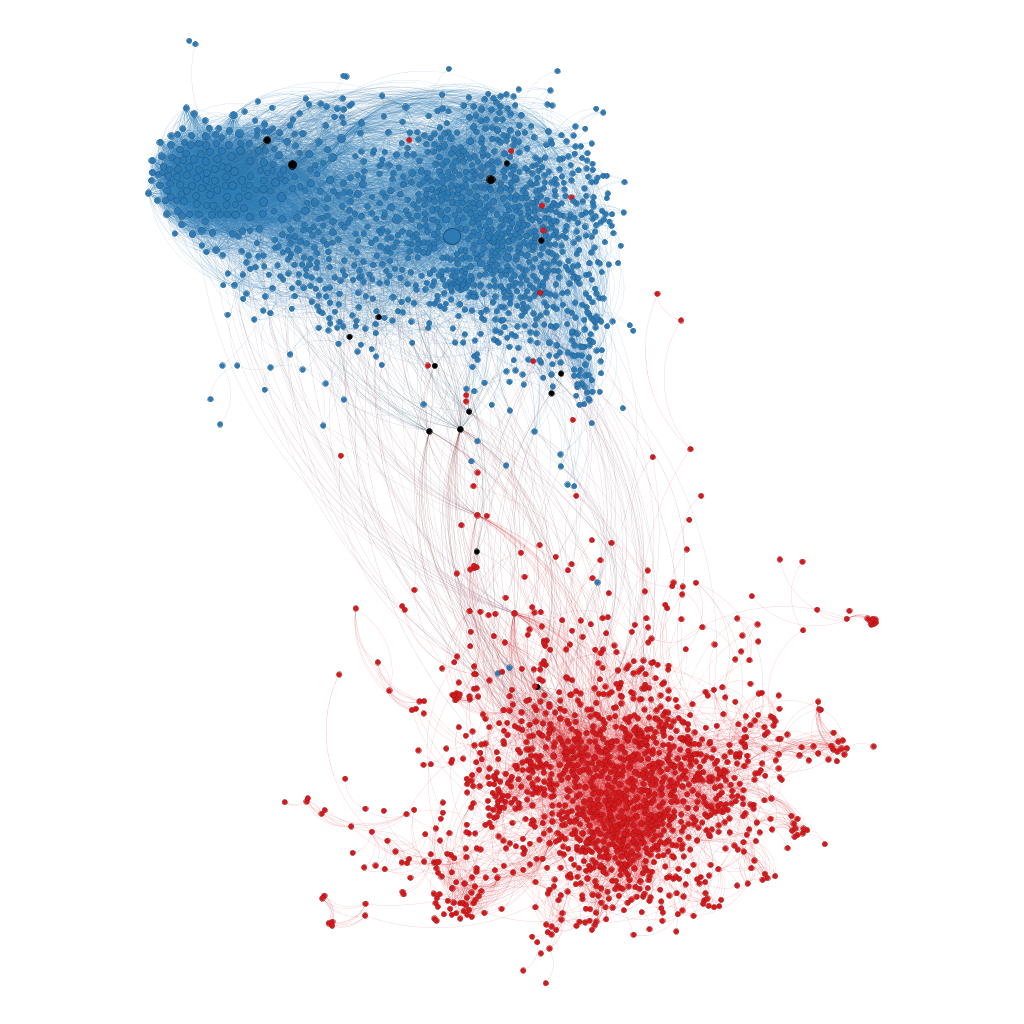}%
        \label{fig:livejournal_gephi}}
    \hfill
    \subfloat[Wiretaping]{%
        \includegraphics[width=0.25\linewidth]{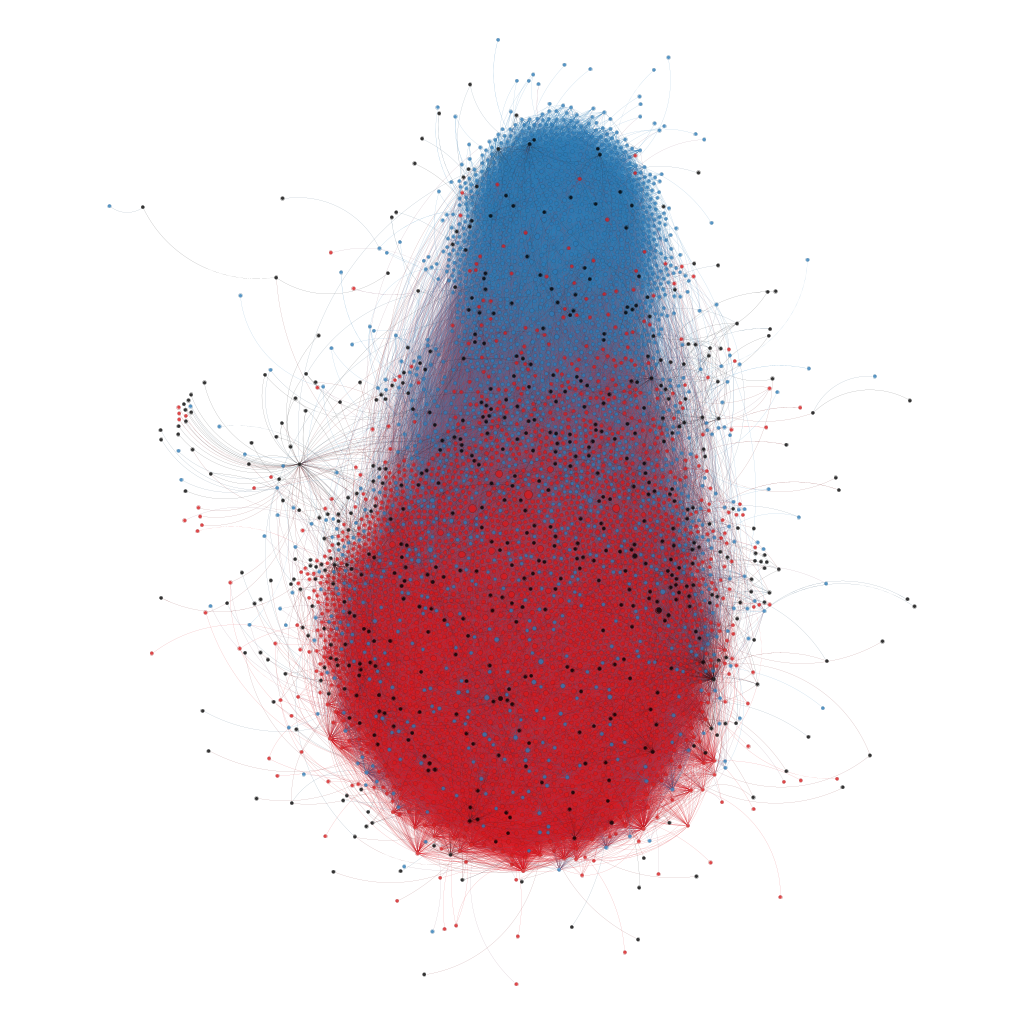}%
        \label{fig:wiretaping}}
    \caption{Node size reflects degree, with larger nodes having higher degrees. Blue and red nodes represent two echo chambers.}
    \label{fig:combined}
\end{figure}

The Political Books network has 105 nodes and 441 edges, with a polarization value of $\pi=0.107$. For the other two real-world networks, LiveJournal and Twitter Wiretapping Scandal, detailed descriptions are provided in Appendix A. However, the key structural properties of all three networks are summarized in Table~\ref{table:1} and a visualization is provided in Fig~\ref{fig:livejournal_gephi} and Fig~\ref{fig:wiretaping}.

\begin{figure*}[t]
    \centering
    \subfloat[Books graph]{%
        \includegraphics[width=0.25\textwidth]{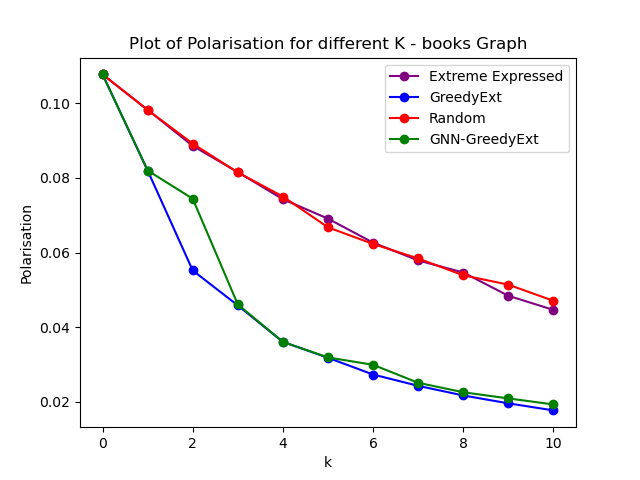}%
        \label{fig:sub1}}
    \hfill
    \subfloat[Livejournal graph]{%
        \includegraphics[width=0.25\textwidth]{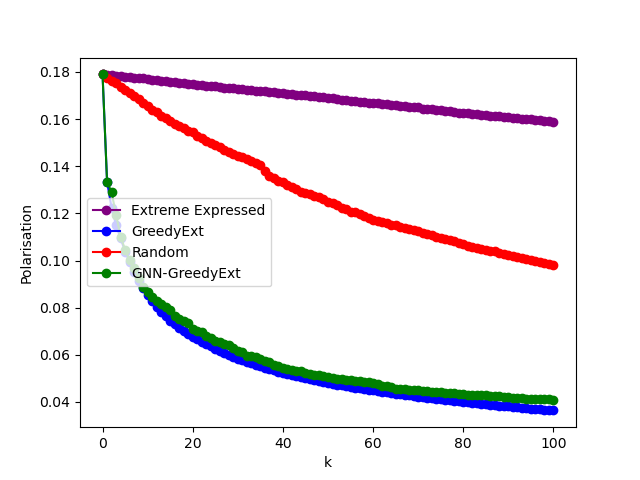}%
        \label{fig:sub2}}
    \hfill
    \subfloat[Wiretaping graph]{%
        \includegraphics[width=0.25\textwidth]{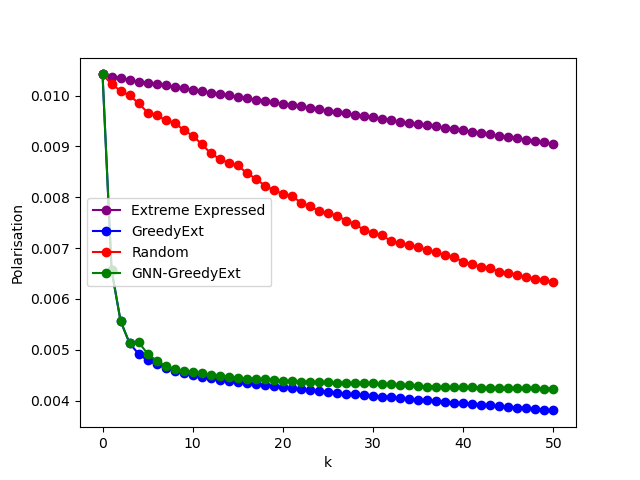}%
        \label{fig:sub3}}
    \caption{Polarization decrease per algorithm timestep}
    \label{fig:combined}
\end{figure*}

\begin{figure}[b]
\begin{center}
\centerline{\includegraphics[width=0.5\columnwidth]{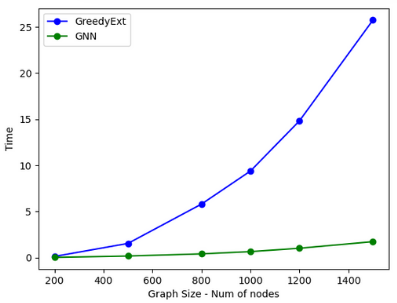}}
\caption{Required time per algorithm for graphs with various sizes}
\label{fig:time_vs_graph}
\end{center}
\end{figure}

\section{Experiments}
\label{sec:5}

We now present the experimental results of our algorithm and compare them against the \textbf{GreedyExt} method. For all experiments, we select small values of $K$ (10\% of the graph size). Following \cite{matakos2017measuring}, we also implement some heuristics to compare our algorithms against.

\subsection{Heuristics}

\paragraph{Random} 
To ensure that the problem instances are non-trivial, we also consider a random \emph{baseline} that selects $K$ nodes to moderate uniformly at random. Cases where the random baseline performs comparably to \textbf{GreedyExt} indicate either (i) insufficient social structure in the network, with nodes being approximately equivalent in influence, and/or (ii) values of $K$ that are too large for the instance under consideration.

\paragraph{ExtremeExpressed} 
An iterative algorithm that, at each step, selects the node with the most extreme expressed opinion, i.e., the node $v$ with the largest $|z_v|$.


\subsection{Depolarization Achieved}

In this first set of experiments, we use our real datasets to demonstrate these two key points: (i) that the problem is not trivial, even in a small scenario, and (ii) that our scheme's performance is close to the ``optimal'' GreedyExt one and outperforms reasonable heuristics such as \textit{ExtremeExpressed}.

In Fig.~\ref{fig:combined} we show the final polarization index of \eqref{eqn:PolarizationIndex}, on the y axis, as the number of nodes $K$ to moderate increases, for four algorithms: (i) GreedyExt, (ii) GNN-GreedyExt, (iii) the random baseline and (iv) the \textit{ExtremeExpressed} heuristic.


\emph{Key Observations}

\begin{enumerate}
    \item The \textbf{GNN-GreedyExt} algorithm follows \textbf{GreedyExt} closely, which means that it makes equivalent choices at each timestep.
    \item The random baseline algorithm fails to decrease the polarization as much as the other algorithms, a fact which implies that the problem is not trivial
\end{enumerate}

According to this figure, the \textbf{GNN-GreedyExt} achieves similar polarization to \textbf{GreedyExt} in all networks. Although \textit{ExtremeExpressed} is a reasonable heuristic we observe that it performs poorly, which is another indicator that the problem is not trivial.

\subsection*{Inference speed}

So far, we have only discussed how the two algorithms compare, in terms of the final polarization that they achieve. As we have already mentioned, the \textbf{GreedyExt} algorithm, is not efficient for large graphs, because of its high computational complexity. Particularly, the evaluation of every node independently at each timestep increases the complexity dramatically. This is exactly where we expect the \textbf{GNN-GreedyExt} to excel, achieving the demonstrated close to optimal depolarization much faster. In the following set of experiments we demonstrate two key points: (i) the \textbf{GNN-GreedyExt} outperforms the \textbf{GreedyExt} in terms of inference speed, and (ii) the inference speed advantage of \textbf{GNN-GreedyExt} increases as the depolarizing graph's size increases.

We compared the two algorithms over the LiveJournal and Wiretaping datasets. Table \ref{table:2} shows the time required for each algorithm to terminate. For this experiment we set $K=50$ without loss of generality. 


\begin{table}[t]
\caption{Time per algorithm in seconds}
\centering
\begin{tabular}{ | c | c | c | } 
  \hline
  Graph & GreedyExt & GNN-GreedyExt  \\ 
   \hline\hline
  LiveJournal & 297  & 31  \\ 
  \hline
  WireTaping & 2571  & 154  \\ 
  \hline
\end{tabular}
\label{table:2}
\end{table}


Clearly, \textbf{GNN-GreedyExt} outperforms the \textbf{GreedyExt} in terms of time. For the Wiretaping graph it terminates 16 times faster. 

Finally, the next experiment compares the time to complete of the two algorithms for graphs with various sizes. In Fig.~\ref{fig:time_vs_graph} the y-axis shows the required time to complete, while the x-axis has the graphs' sizes. As we can see, the required time for the \textbf{GreedyExt} algorithm grows very quickly with respect to the graph size. In contrast the time to complete for the \textbf{GNN-GreedyExt} grows much slower, with respect to the size of the graph.

Moreover, the time advantage of the \textbf{GNN-GreedyExt} against the \textbf{GreedyExt}, is larger, for larger graphs. In other words, it increases with respect to the size of the underlying graph. In the examples that we have shown where large graphs were involved, the LiveJournal and the Wiretaping datasets, the \textbf{GNN-GreedyExt} is $\times 10$ and $\times 16$ times faster respectively.

\section{Conclusion}
\label{sec:6}

In this work, we proposed an alternative approach to address the \textbf{ModerateExpressed} problem, originally introduced in \cite{matakos2017measuring}. One of the key limitations of their solution is its lack of scalability, as their algorithm becomes increasingly slow, when applied to larger graphs. Our method leverages GNNs, specifically a 2-layer GCN, to generate node embeddings. Instead of using brute force evaluation, we apply a simple argmax function to identify the node with the highest score. In terms of polarization, our approach achieves results almost identical to those of the original algorithm. However, the main advantage of our method is its ability to scale efficiently to larger graphs due to the GNN architecture. Our results show that the \textbf{GNN-GreedyExt} dramatically improves the required time to solve the problem and thus it is a great choice when dealing with large graphs. An interesting future direction is to explore the submodularity of the problem \textbf{ModerateExpressed} and quantify the suboptimality for greedy algorithms like the ones we have discussed.

\begin{acks}
This project has received funding from the European Union’s Horizon Europe research and innovation programme under the Marie Skłodowska-Curie (MSCA) grant agreement No 101226860 (project GENOME) and
from the Hellenic Foundation for Research \& Innovation (HFRI) project 8017: “AI4RecNets: Artificial Intelligence (AI) Driven Co-design
of Recommendation and Networking Algorithms”.
\end{acks}

\bibliographystyle{ACM-Reference-Format}
\bibliography{sample-base}

\clearpage


\appendix
\section{Real world networks}

\subsection{LiveJournal}

LiveJournal is a free online blogging community where users can declare friendships and form user-defined groups, which serve as ground-truth communities. This dataset was introduced by Yang and Leskovec \cite{yang2012defining} and is publicly available through the SNAP repository. The dataset contains multiple communities; however, our study requires a network with a dual echo chamber structure.

To construct such a network, we began with the 5000 highest-quality communities identified in the original study \cite{yang2012defining}, where “quality” is formally defined. These communities were then ranked by size, and the two largest were selected to form the graph. The resulting network, shown in Fig.~\ref{fig:livejournal_gephi}, exhibits a clear dual echo chamber structure. Node opinions are assigned based on community membership.

\subsection{Twitter Wiretapping Scandal}

This dataset captures political discussions on Twitter regarding the Greek Wiretapping Scandal, which was revealed in 2022. A detailed description is provided in \cite{dimitriadis2024analyzing}. In brief, the dataset contains tweets with hashtags and keywords relevant to the scandal, from which we construct the retweet graph.  

User opinions are inferred using a heuristic based on their following behavior. Specifically, for each user we count the number of politicians followed, and assign the user’s supported party as the one from which they follow the most politicians. In case of a tie, the user is assumed to have no party affiliation. The opinion variable is then set to $s_i = +1$ if the supported party belongs to the left of the political spectrum and $s_i = -1$ otherwise.  

The complete retweet graph contains approximately $44{,}000$ nodes. For our experiments, we extract a representative sample of $5{,}000$ nodes, shown in Fig.~\ref{fig:wiretaping}. Owing to its relatively high density, this network exhibits lower polarization compared to the other graphs considered.









\end{document}